\documentclass[10pt,twocolumn,letterpaper]{article}            
\usepackage[pagenumbers]{wacv} 

\usepackage{microtype}
\usepackage{graphicx}
\usepackage{booktabs} 
\usepackage{nicefrac}
\usepackage{booktabs, longtable, multirow}
\usepackage{amsmath}
\usepackage{amssymb}
\usepackage{soul}
\usepackage{comment}
\usepackage{mathtools}
\usepackage{amsthm}
\usepackage{adjustbox}
\usepackage{nicefrac}

\newcommand{\R}[1]{\mathbb{R}^{#1}}

\definecolor{annotation}{RGB}{0, 0, 0}

\usepackage[accsupp]{axessibility}
\usepackage{svg}
\definecolor{wacvblue}{rgb}{0.21,0.49,0.74}
\usepackage[pagebackref,breaklinks,colorlinks,allcolors=wacvblue]{hyperref}

\usepackage{xspace}

\let\OldS\S
\renewcommand{\S}{\OldS\xspace}

\title{HodgeFormer: Transformers for Learnable Operators on Triangular Meshes through Data-Driven Hodge Matrices}

\author{Akis Nousias\\
\small K3Y Labs
\\{\tt\small anousias@k3y.bg}
\and
Stavros Nousias\\
 \small Technical University of Munich
\\{\tt\small stavros.nousias@tum.de}
}

\begin{document}
\maketitle
\begin{abstract}
Currently, prominent Transformer architectures applied to graphs and meshes for shape analysis tasks employ traditional attention layers that heavily utilize spectral features requiring costly eigenvalue decomposition-based methods. 
To encode the mesh structure, these methods derive positional embeddings that heavily rely on eigenvalue decomposition-based operations, e.g. on the Laplacian matrix, or on heat-kernel signatures, which are then concatenated to the input features.
This paper proposes a novel approach inspired by the explicit construction of the Hodge Laplacian operator in Discrete Exterior Calculus as a product of discrete Hodge operators and exterior derivatives, i.e. $(L := \star_0^{-1} d_0^T \star_1 d_0)$. We adjust the Transformer architecture in a novel deep learning layer that utilizes the multi-head attention mechanism to approximate Hodge matrices $\star_0$, $\star_1$, and $\star_2$ and learn families of discrete operators $L$ that act on mesh vertices, edges and faces. Our approach results in a computationally-efficient architecture that achieves comparable performance in mesh segmentation and classification tasks, through a direct learning framework, while eliminating the need for costly eigenvalue decomposition operations or complex preprocessing operations.
\textit{Code and data are publicly available at} \url{https://github.com/hodgeformer/}
\end{abstract}

\section{Introduction}
\label{sec:intro}

Data-driven algorithms are at the core of machine learning innovations in fields such as natural language processing, computer vision, and audio processing. Over the past decade, existing architectures and handcrafted features have been gradually replaced by machine learning models that are able to capture fine-grained data-driven structures and patterns, better leverage the underlying hardware, and can efficiently handle large-scale data.

In recent years, Transformer-based architectures have been at the forefront of breakthroughs in all of the aforementioned fields, due to their scalability in terms of data and hardware, their generality as an architecture, and their versatility in accommodating different data modalities \cite{khan2022transformers}.

Traditional Transformer-based approaches for 3D data typically follow a graph-based application paradigm. Vanilla Transformers are used, usually with node features as input, while structural information is incorporated through positional embeddings, usually via eigenvalue decomposition of a Laplacian-based matrix \cite{vecchio2023met,lin2021mesh,li2022meshformer}.

In this work, we introduce a novel Transformer architecture inspired by the explicit construction of the Laplacian operator in Discrete Exterior Calculus (DEC) adjusted to triangular 3-dimensional meshes. This kind of connection is not novel by itself and was the main idea of HodgeNet by \citet{smirnov2021hodgenet}, where input features were used for learning Hodge Star operators as diagonal matrices. These matrices were then used for constructing Laplacian operators, which were then applied to input features in a spectral learning kind of approach.
Specifically, we draw connections between discrete Hodge Star operators and Transformer-style attention mechanisms. While most common constructions of discrete Hodge Star operators are limited to diagonal matrices, these represent only some specific Hodge operator realizations. In fact, the Galerkin method, widely used in Finite Element Methods (FEM) literature \cite{suli2000hp}, prescribes systematic ways to discretize various differential operators, including the Hodge operator. The key idea is to project onto carefully designed basis functions (e.g., piecewise linear or higher order elements) and account for the overlap of these basis functions. These overlaps are used to design the mass matrix, which acts as a discrete Hodge operator. Unlike the diagonal Hodge operator, which considers only self-contributions, the Galerkin Hodge operator \cite{auchmann2006geometrically,mohamed2016comparison,lim2020hodge} incorporates contributions from neighboring elements, resulting in a sparse, non-diagonal matrix.

There are several key insights that allow drawing direct connections with the Transformer-style attention. Specifically, the projection with matrices $W_Q$, $W_K$, $W_V$ onto a learned space corresponds to a basis projection as in the Galerkin method, and the self-attention mechanism $Q \cdot K^T$ encodes pairwise interactions and accounts for overlaps of the basis functions. If the self-attention mechanism is localized, then it aligns with Galerkin-style discretizations \cite{cao2021choose}.
The proposed architecture includes a novel Transformer-inspired layer that enables information propagation directly on the manifold through attention-based learnable Hodge Star matrices and incidence matrices, which serve as discrete exterior derivatives.
By incorporating the structure considerations into the architecture, we do not rely at all on eigen-decomposition based methods, spectral features, or preprocessing steps, while achieving competitive performance compared to state-of-the-art approaches on meshes.

The rest of the paper is organized as follows: \cref{sec:rw} presents the related work. \cref{sec:preliminaries} provides preliminaries on aspects relevant to the proposed approach.  \cref{sec:hodge_attention} presents the methodology, and \cref{sec:evaluation} presents the experimental results. Finally, \cref{sec:conclusion} discusses the outcomes of this work and concludes this paper.
\section{Related Work}
\label{sec:rw}

\noindent \textbf{Mesh convolutions.} Recent advances in geometric deep learning have adapted convolution operations for sampled 3D manifolds by exploiting relationships between mesh elements, \ie{} vertices, edges and faces, to construct convolutional filters. MeshCNN, introduced by Hanocka et al. \cite{hanocka2019meshcnn}, proposed edge-based mesh convolution, mesh pooling and mesh unpooling operations. However, the method was limited by the fixed topology requirement and potential loss of geometric information during pooling. DualConvMesh-Net \cite{schult2020dualconvmesh} attempted to bridge local and global features through geodesic and Euclidean convolutions, but faces computational challenges with large meshes.
Another category of methods utilizes subdivision schemes to enable convolution operations on meshes. SubdivNet \cite{de2016subdivision} and Subdivision-based Mesh Convolutional Networks \cite{hu2022subdivision} enforce regular structure through remeshing or subdivision techniques. The latter defines convolution operations directly on 3D triangle meshes via subdivision structure, treating mesh faces as fundamental convolution units analogous to 2D pixels. However, these subdivision-driven approaches suffer from several limitations, potentially introducing geometric artifacts, failing to preserve fine local features, requiring specific connectivity patterns, such as loop subdivision sequences, and incurring high computational costs from remeshing operations.

\noindent \textbf{Spectral methods} are based on the eigenvalue decomposition of discrete operators, typically the Laplace-Beltrami operator, utilizing the translation, rotation, and scale invariance of operations in the spectral domain.
HodgeNet \cite{smirnov2021hodgenet} introduces a learnable class of sparse operators on meshes, built from standard constructions in discrete exterior calculus. It employs standard geometry processing operations and Hodge star operators ($\star_0$ and $\star_1$) to encode per-vertex and per-edge features.
Spectral approaches by \cite{defferrard2016convolutional} leverage the graph Laplacian for convolution, with subsequent works like CayleyNets \cite{levie2018cayleynets} improving spectral filtering design. However, these methods struggle with computational complexity due to the high computational cost of eigenvalue decomposition.
To avoid this limitation, Surface Networks \cite{kostrikov2018surface} leverage Graph Neural Networks (GNNs) to learn polynomial expansions of both intrinsic (Laplacian) and extrinsic (Dirac) operators to capture mesh geometry, enabling them to process irregular mesh structures directly.
Shape descriptors built on the Laplace-Beltrami operator have been particularly successful in capturing geometric properties. The Heat Kernel Signature (HKS)~\cite{sun2009concise} captures multiscale geometric information through the heat diffusion process, providing an isometry-invariant point descriptor. The Wave Kernel Signature (WKS)~\cite{aubry2011wave} offers an alternative based on quantum mechanical principles, achieving better feature localization. 
Building on these descriptors, several neural architectures have emerged. DiffusionNet~\cite{sharp2020diffusion} leverages diffusion-based features to achieve robust segmentation and classification on partial and non-isometric shapes. Geodesic CNNs~\cite{masci2015geodesic} introduce geometry-adapted convolutional and pooling layers that operate in the spectral domain, utilizing HKS and WKS features as input channels for learning intrinsic shape representations.
LaplacianNet \cite{qiao1910laplaciannet} introduces a deep learning framework for 3D meshes encoding mesh connectivity using Laplacian spectral analysis. The network uses Mesh Pooling Blocks (MPBs) to split the mesh surface into local pooling patches and aggregate both local and global features. A mesh hierarchy is built from fine to coarse using Laplacian spectral clustering, making the method robust to different triangulations and isometric transformations. The method depends on eigenvalue decomposition and spectral clustering.


\noindent \textbf{Graph Learning-based} E(n)-Equivariant Graph Neural Networks (EGNN) \cite{satorras2021n} efficiently extend message passing to geometric graphs by maintaining scalar invariant features and equivariant coordinates for each node, using relative distances to preserve rotations, translations, and reflections. Equivariant Mesh Neural Networks (EMNN)\cite{trang20243} build on this by adding surface-aware messages that leverage mesh face information. EMNN further incorporates multiple vector channels and hierarchical pooling to capture long-range dependencies while maintaining E(3)-equivariance.

\noindent \textbf{Transformer adaptations} display potential in capturing global relationships in 3D meshes. The Laplacian Mesh Transformer~\cite{li2022laplacian} applies Transformer layers to Laplacian eigenvectors and geometric features. MeshMAE~\cite{liang2022meshmae} introduces self-supervised learning through masked autoencoders for mesh understanding. MeshFormer~\cite{liu2024meshformer} addresses scalability by combining mesh simplification with hierarchical graph-based attention mechanisms. Mesh Transformer (MeT)~\cite{vecchio2023met} proposes a two-stream architecture that processes triangle and cluster features in parallel, using Laplacian eigenvectors for positional encoding.
Despite their effectiveness, these Transformer-based methods face significant computational challenges. The eigenvalue decomposition required for Laplacian features and the quadratic complexity of attention mechanisms limit their applicability to large-scale meshes, while the need for complex architectural components increases implementation and training difficulty.
\section{Preliminaries}
\label{sec:preliminaries}

\subsection{Triangular meshes \& Discrete Operators}
In this work, we focus on triangular meshes $\mathcal{M(V, E, F)}$ with vertices $n_v = |\mathcal{V}|$, $n_e = |\mathcal{E}|$ edges, and $n_f = |\mathcal{F}|$ faces. 
In the context of Discrete Exterior Calculus (DEC) \cite{desbrun2005discrete}, a triangular mesh is seen as a 2-dimensional oriented simplicial complex. It is composed of oriented $k$-simplices for $k \in \{0, 1, 2\}$ corresponding respectively to vertices, edges, and faces. On each set of $k$-simplices, we consider discrete differential $k$-forms $\Omega^{k}$, that is, functions that assign values to the elements of each set. In addition, we employ two important constructs:


\begin{enumerate}[i]
   \item The \textit{discrete exterior derivative} $d_k$ along with its adjoint $d_k^T$, that is, the linear map $d_k: \Omega^{k} \rightarrow \Omega^{k+1}$ mapping discrete $k$-forms to discrete $k+1$ forms. 
   
   \item The \textit{discrete Hodge Star}, which is the discrete analog of the Hodge Star operator $\star_k: \Omega^{k} \rightarrow \widetilde{\Omega}^{n-k}$ that maps k-forms $\Omega^{k}$ to their dual $(n-k)$ forms $\widetilde{\Omega}^{n-k}$.
\end{enumerate}
The discrete exterior derivative and the discrete Hodge Star operator encode different aspects of the mesh. The exterior derivative captures the orientation and connectivity between mesh elements, translating how vertices, edges, and faces relate, while the Hodge Star operator encodes metric information such as angles, lengths, areas, and volumes.

A key feature of the discrete Hodge Star, arising from the nature of the discretization, is that it does not need to act on forms defined on the same mesh. Instead, it naturally relates quantities on two distinct and possibly unrelated meshes: the \textit{primal mesh} (the original simplicial complex) and the \textit{dual} (or secondary) mesh. This decoupling offers a wide range of choices for the dual mesh, and indeed, the literature presents many dual mesh discretizations, such as the circumcentric dual, the barycentric dual, and others that lead to different Hodge Star instantiations. For a triangular mesh, the corresponding sets of primal and dual elements share the same dimensionality, although this correspondence does not hold for other simplicial complexes.

\begin{figure}[ht]
    \begin{center}  \centerline{\includegraphics[width=0.7\linewidth]{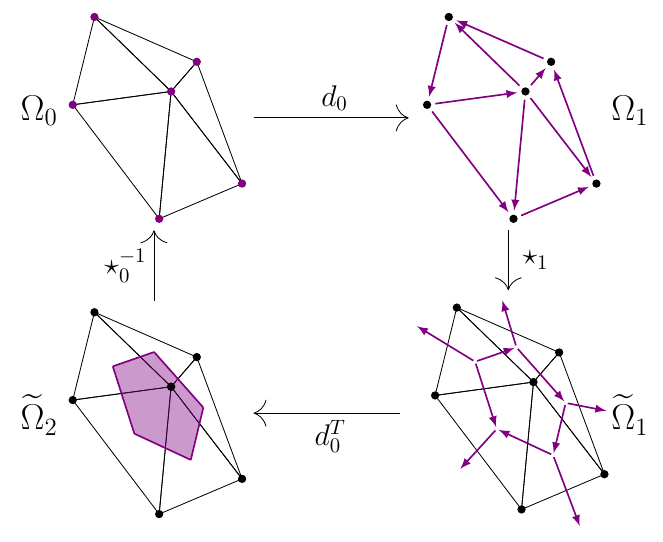}}
        \caption{The Hodge Laplacian applied on a 0-form, i.e. values on vertices, as a sequence of discrete exterior derivative and Hodge Star operators.}     
        \label{fig:laplacian_on_vertices}
    \end{center}
    \vspace{-3em}
\end{figure}

All the above constructs can be naturally represented as matrices, as shown in  \cref{dec-objects-to-matrices}. The matrix representations of the discrete exterior derivative operators $d_0 \in \{-1, 0, 1\}^{n_e \times n_v}$ and $d_1 \in \{-1, 0, 1\}^{n_f \times n_e}$ are simply the \textit{signed incidence matrices} of the input meshes. The matrix representations of the Hodge Star operators, as mentioned above, depend on the discretization of the dual mesh.

\begin{table}[!htbp]
    \centering
    \begin{tabular}{@{}p{4cm}p{3.9cm}@{}}
\toprule
    \textbf{Object} & \textbf{Matrix Representation}\\ 
\midrule
    $\Omega^{0}$ (0-form)      &  $x_v \in \mathbb{R}^{n_v\times d_v}$ \\
    $\Omega^{1}$ (1-form)      &  $x_e \in \mathbb{R}^{n_e\times d_e}$ \\
    $\Omega^{2}$ (2-form)      &  $x_f \in \mathbb{R}^{n_f\times d_f}$ \\ 
    $d_{0}$ (vertices-to-edges)        &    $d_0 \in \{-1, 0, 1\}^{n_e \times n_v}$\\
    $d_{1}$ (edges-to-faces)           &    $d_1 \in \{-1, 0, 1\}^{n_f \times n_e}$\\
    $\star_0$ (Hodge Star on 0-forms)  &    $\star_0 \in \mathbb{R}^{n_v\times n_v}$\\
    $\star_1$ (Hodge Star on 1-forms)  &    $\star_1 \in \mathbb{R}^{n_e\times n_e}$\\
    $\star_2$ (Hodge Star on 2-forms)  &    $\star_2 \in \mathbb{R}^{n_f\times n_f}$\\
\bottomrule 
\end{tabular}
    \caption{Mapping of different objects to matrices}
    \label{dec-objects-to-matrices}
    \vspace{-1em}
\end{table}

\noindent By composing discrete operators, we can construct complex objects such as the curl operator ($\star_2 \cdot d_1$), the divergence operator ($\star_0^{-1} d_0^T \star_1$), or the Laplacian operator on vertices $L_v := \star_0^{-1} d_0^T \star_1 d_0$, as shown in Figure \ref{fig:laplacian_on_vertices}, or the more general Hodge Laplacian on k-forms (also known as k-form Laplacian, or de Rham Laplacian):
\begin{equation}
    L_k := d_{k-1} \cdot \star_{k-1}^{-1} \cdot d_{k-1}^{T} \cdot \star_{k} + \\
          \star_{k}^{-1} \cdot d_{k}^{T} \cdot \star_{k+1} \cdot d_{k}
    \label{eq:hodge-laplacian}
\end{equation}

\subsection{Transformers}
A Transformer is a function 
$
T : \mathbb{R}^{n \times d_m} \to \mathbb{R}^{n \times d_{m}}
$
where \( d_{m} \) is the embedding (model) dimension, defined by the composition of \( L \) Transformer layers \( T_1(\cdot), \dots, T_L(\cdot) \):

\begin{equation}
   T_l(x) = f_l(A_l(x) + x).
   \label{eq:transformer}
\end{equation}

Here, \( f_l(\cdot) \) transforms each row of shape \( \mathbb{R}^{1 \times d_m} \) independently of the others and is usually implemented with a small two-layer feedforward network. \( A_l(\cdot) \) is the self-attention function and is the only component acting across sequence positions. The self-attention function \( A_l(\cdot) \) computes, for each position, a weighted sum of all positions in the sequence. Formally, for input \( x \in \mathbb{R}^{n \times d_{m}} \):
\begin{equation}
Q = x W_Q, \quad K = x W_K, \quad V = x W_V
\label{eq:attn_qkv}
\end{equation}
Typically \( W_Q, W_K, W_V \in \mathbb{R}^{d_{m} \times d_h} \) where $d_h = d_{m}/{h}$, with $h$ the number of attention heads, and \( Q \in \mathbb{R}^{n \times d_h} \), \( K \in \mathbb{R}^{n \times d_h} \), \( V \in \mathbb{R}^{n \times d_h} \). The attention output is:
\begin{equation}
\text{Attention}(Q, K, V) = \text{softmax}\left( \frac{Q K^\top}{\sqrt{d_h}} \right) V
\label{eq:attn}
\end{equation}
and the output shape is \( \mathbb{R}^{n \times d_h} \).
For multi-head attention, this process is repeated in parallel for each head; the outputs are concatenated and projected back to $d_{m}$. The feedforward network is applied independently to each position (row) and is typically a two-layer MLP.

\begin{figure}[!t]
    \begin{center}
        \centerline{\includegraphics[width=\linewidth]{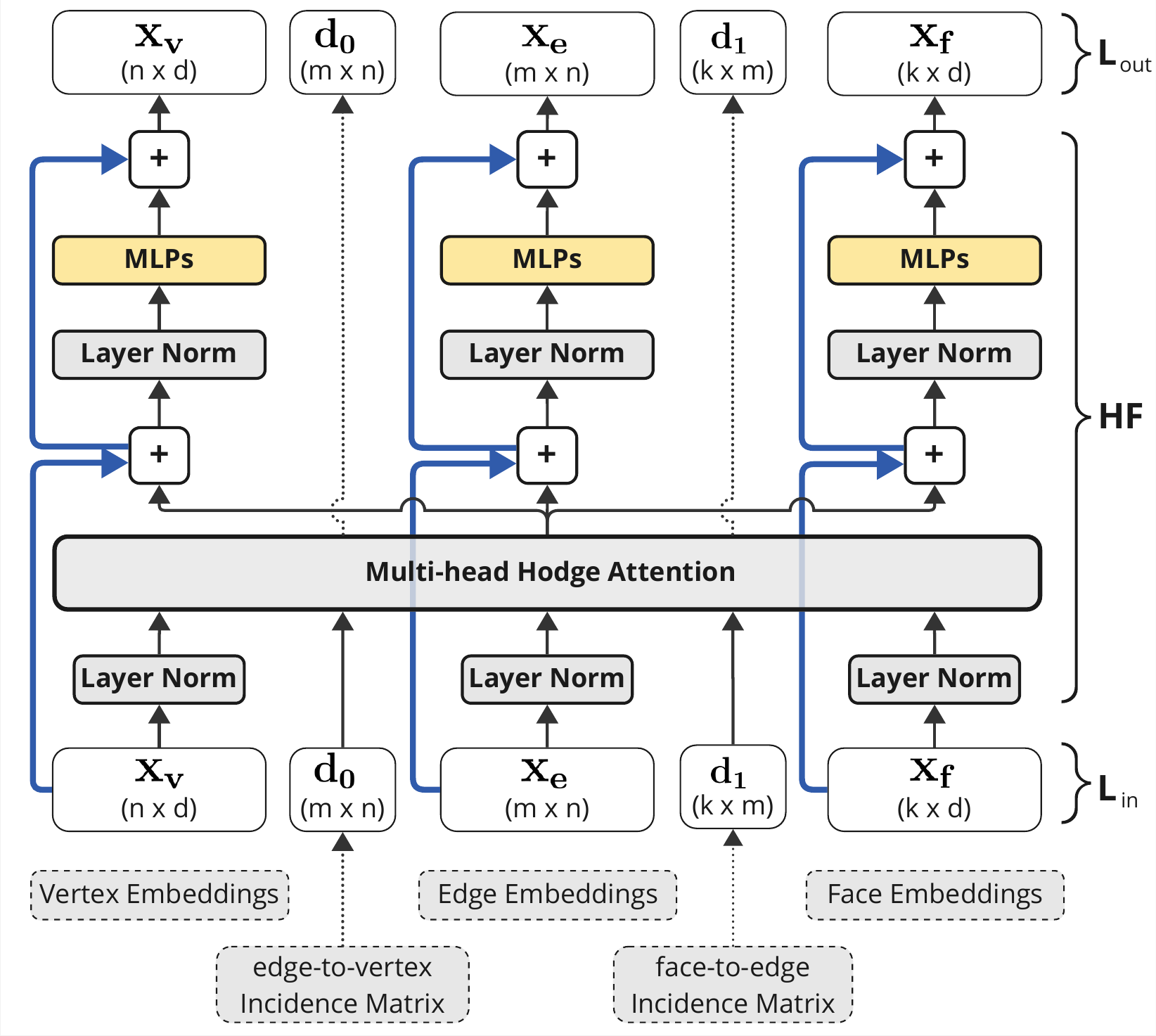}}
        \caption{Overview of a HodgeFormer layer operating on vertex, edge and face features via the Multi-Head Hodge Attention mechanism. The layer can be configured so that different mesh elements or combinations of them will be updated.}     
        \label{fig:architecture_hodge_attention}
    \end{center}
    \vspace{-3em}
\end{figure}

\begin{figure}[!ht]
    \begin{center}
    \centerline{\includegraphics[width=0.96\linewidth]{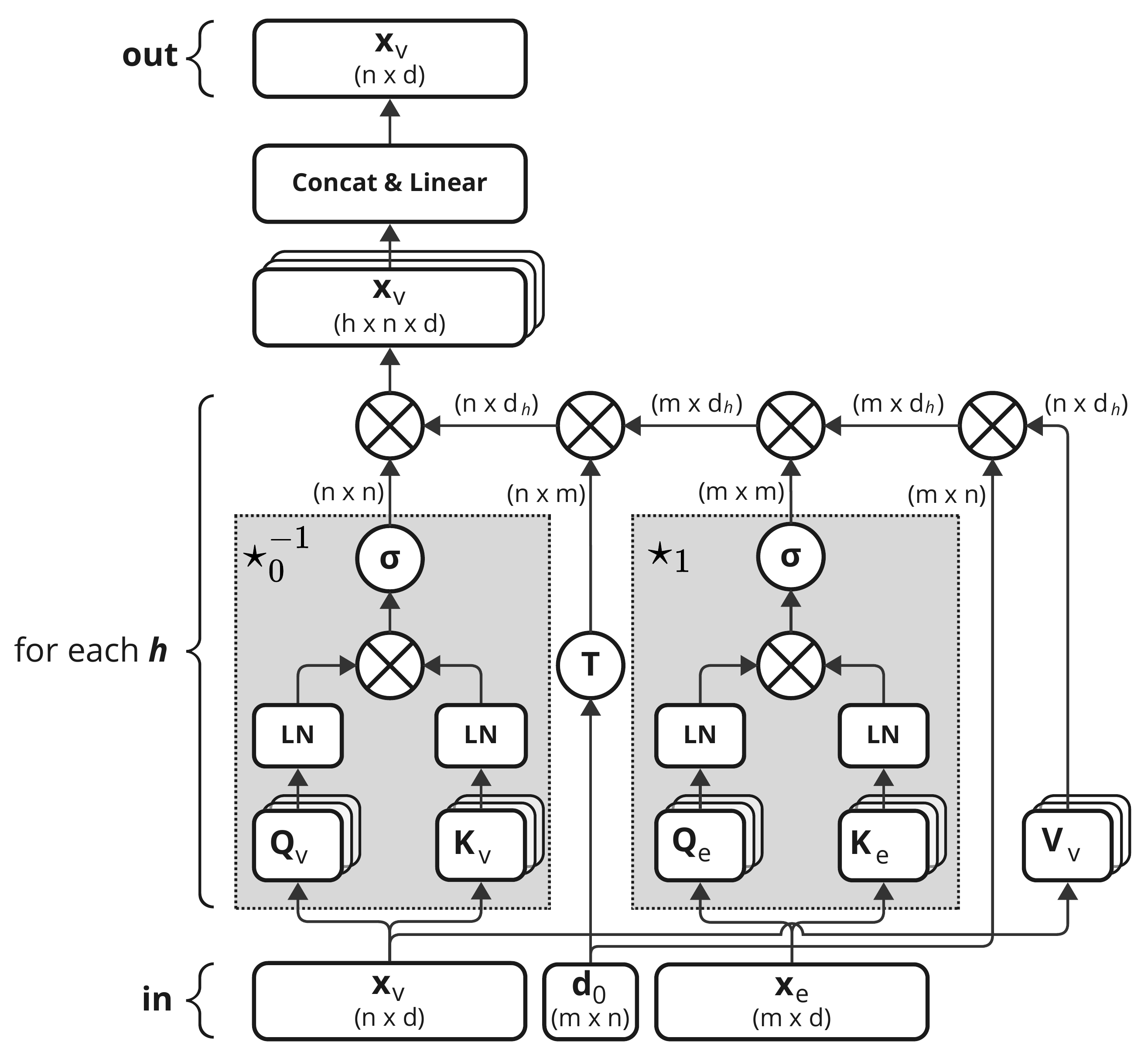}}
        \caption{Multi-head Hodge Attention applied on latent vertex features $x_v$. The multi-head attention mechanism learns data-driven Hodge Star matrices $\star_0^{-1}$ and $\star_1$. \textcolor{annotation}{Layer Norm (LN) is applied on $Q$, $K$ matrices to increase training stability.}}
        \label{fig:formation_xv}
    \end{center}
    \vspace{-3em}
\end{figure}

\section{A Transformer Model with Hodge Attention}
\label{sec:hodge_attention}

\subsection{HodgeFormer Layer}

We propose a novel deep learning architecture shown in \cref{fig:architecture_hodge_attention}, named ``HodgeFormer'', that leverages the multi-head attention mechanism for learning data-driven Hodge Laplacian operators on $k$-forms based on the parametrization of the Hodge Star $\star_k$. Let $x_k \in \mathbb{R}^{n_k\times d_k}$ be a $k$-form on mesh elements with $k \in \{v, e, f\}$, respectively, of size $n_k$ and let $W_{Q_k} \in \R{d_k \times d_{h_k}}$, $W_{K_k} \in \R{d_k \times d_{h_k}}$, and $W_{V_k} \in \R{d_k \times d_{h_k}}$ the learnable projection matrices transforming $x_k$ to corresponding representations $Q_k$, $K_k$, and $V_k$ as in \cref{eq:attn}. For simplicity, we set $d_k = d$ for $k \in \{v, e, f\}$, and $d_{h_k} = d_h = d / h$ for $h$ attention heads.
\noindent Then we can parametrize the Hodge Star operator $\star_k$ as:
\begin{align}
    \star_{k}(x_k) &= \sigma\left({\frac{Q_k K_{k}^T}{\sqrt{d_h}}}\right),      & \text{for }  k \in \{v, e, f\}
    \label{eq:hodge-star-matrix}
\end{align}
where $\sigma\left(\cdot\right)$ is the row-wise softmax non-linearity. For the inverse Hodge Star operators, i.e., $\star_{k}^{-1}$, we employ dedicated linear maps ${Q_k},{K_k}$.


\noindent We define the HodgeFormer layer $H_{l}$ as the function that operates and updates k-forms on vertices, edges and faces:
\begin{equation}
    H_l(x_v, x_e, x_f) := G_l(A_{H_{l}}(x_v, x_e, x_f) + (x_v, x_e, x_f))
\label{eq:HF_layer}
\end{equation}
where $A_{H_l}(\cdot\space, \cdot\space, \cdot)$ is the Multi-head Hodge Attention operating on each k-form:
\begin{align} 
  \label{eq:MHA}
  x_v, x_e, x_f  \enspace :=& \enspace A_{H_l}(x_v, x_e, x_f) \\
  \label{eq:MHA_per_k-form_v}
  x_v = A_v(x_v, x_e) \enspace =&  \enspace L_v(x_v, x_e)      \cdot V_v(x_v) \\
  \label{eq:MHA_per_k-form_e}
  x_e = A_e(x_v, x_e, x_f) \enspace =& \enspace L_e(x_v, x_e, x_f) \cdot V_e(x_e) \\
  \label{eq:MHA_per_k-form_f}
  x_f = A_f(x_e, x_f)  \enspace =& \enspace L_f(x_e, x_f)      \cdot V_f(x_f)
\end{align} 
and $G_l(\cdot)$ a function that transforms the values of each k-form independently and is implemented with separate two-layer feedforward networks with residual connections.

\noindent From \cref{eq:hodge-laplacian}, the Hodge Laplacians $L_v$, $L_e$, $L_f$ are defined for vertices (0-form), edges (1-form) and faces (2-form) as :
\begin{align} 
    \begin{split}
        L_v \enspace := \enspace & \, \star_0^{-1}(x_v) \cdot d_0^T \cdot \star_1(x_e) \cdot d_0 \\
        L_e \enspace := \enspace & \, d_0 \cdot \star_0^{-1}(x_v) \cdot d_0^T \cdot \star_1(x_e) \, + \\ 
        & \quad \, \, \, \,          \star_1^{-1}(x_e) \cdot d_1^T  \cdot \star_2(x_f) \cdot d_1\\
        L_f \enspace := \enspace & \, d_1 \cdot \star_1^{-1}(x_e) \cdot d_1^T \cdot \star_2(x_f)
        \label{eq:learned-operators}
    \end{split}
\end{align}
\textcolor{annotation}{
The HodgeFormer layer $H_l$ extends the standard transformer layer to independently operate and update $k$-form features using multi-head attention via mesh-based Hodge Laplacians and Stars, closely matching the residual and feedforward update principles in canonical transformer networks.
}

\subsection{End-to-end Architecture}
\label{sec:end-to-end-architecture}
\begin{figure}[t]
    \begin{center}    
        \centerline{
        
    \includegraphics[width=\linewidth]{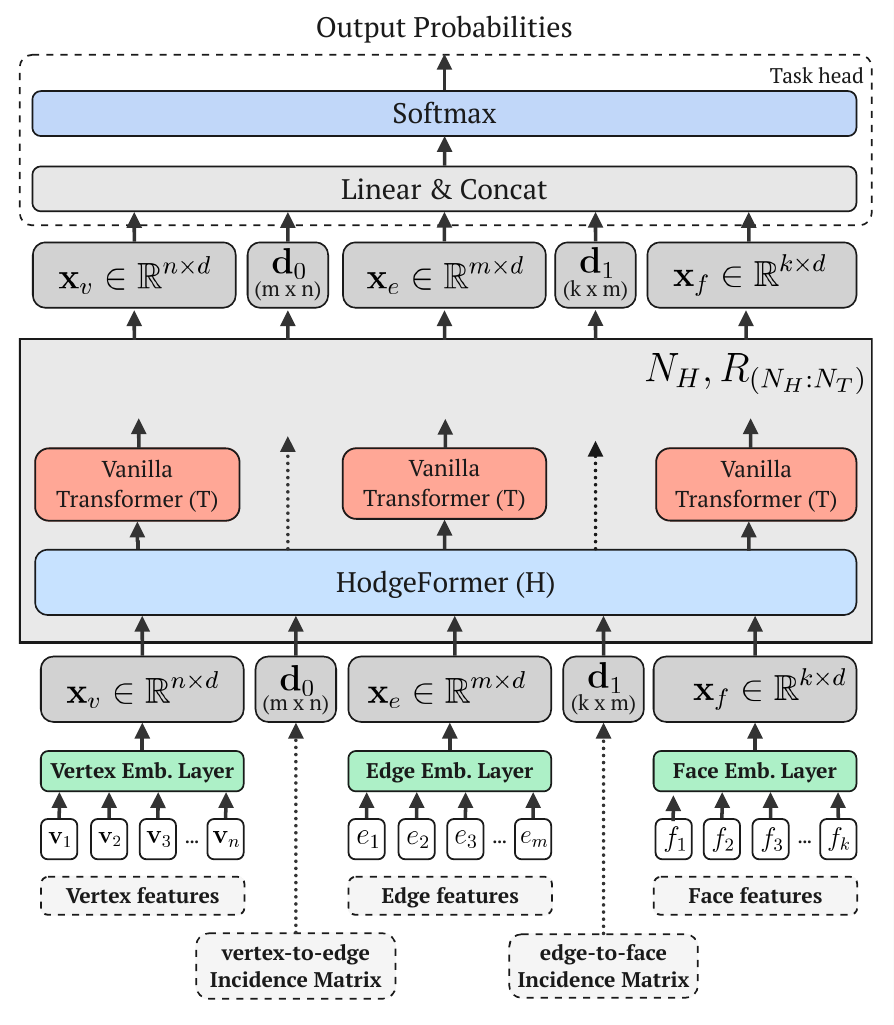}
        }
        
        \caption{Overview of a deep learning architecture with combined HodgeFormer and vanilla Transformer layers operating on an input triangular mesh.}
        \label{fig:overview}
    \end{center}
    \vspace{-3.0em}
\end{figure}

This section presents the proposed end-to-end architecture. As illustrated in \cref{fig:overview}, the architecture consists of (i) Embedding Layers, (ii) HodgeFormer and Vanilla Transformer layers, and (iii) a task-specific head.

The input consists of features extracted from the mesh, along with the sparse oriented incidence matrices $d_0$ and $d_1$. Dedicated embedding layers, \textcolor{annotation}{discussed in \cref{sec:embedding-layer}}, operate separately on vertex, edge, and face features and map them to target latent dimensions $d_v$, $d_e$, and $d_f$. Then, the latent embeddings for each mesh element are transformed by a mixture of sequentially placed HodgeFormer and Transformer layers and updated accordingly. The final embeddings are passed to the task head, which maps them to the task-specific output dimensions. Putting together HodgeFormer layers with vanilla Transformer layers can be thought of as combining local operators with global mixers. Different mixing strategies of HodgeFormer and vanilla Transformer layers can be applied.

The proposed end-to-end architecture is configurable to combine operations on different mesh elements, i.e., vertices, edges, and faces, where each learned operator acts upon and updates specific k-forms, as depicted in \cref{tab:architecture-modes}. In the case that specific mesh elements are not selected to be updated, the corresponding Transformer layers are omitted. The size of the architecture in terms of trainable parameters depends on the number and configuration of used layers and the combination of mesh elements selected to operate upon.

\begin{table}[t]
    \centering
    \begin{tabular}{@{}lclll@{}}
\toprule
    & \textbf{Acts~on} & \textbf{Learns}  & \textbf{Requires} & \textbf{Updates}\\ 
\midrule 
    $L_v$ & $v$ & $\star_0^{-1}$, $\star_1$                            & $x_v$, $x_e$      &  $x_v$ \\
    $L_e$ & $e$ & $\star_0^{-1}$, $\star_1^{-1}$, $\star_1$ $\star_2$  & $x_v$, $x_e$, $x_f$  &  $x_e$ \\
    $L_f$ & $f$ & $\star_1^{-1}$, $\star_2$                            & $x_e$, $x_f$      &  $x_f$ \\ 
\bottomrule 
    \end{tabular}
    \vspace{-0.5em}
    \caption{Overview of learnable operators $L_v$, $L_e$, $L_f$.}
    \label{tab:architecture-modes}
    \vspace{-1.5em}
\end{table}

\subsection{Sparse Attention}

For the multi-head Hodge attention mechanism, we employ sparse attention by defining sparsity patterns based on local neighborhoods of mesh elements. Instead of computing the full attention matrix, this choice reduces complexity of the attention operator and aligns with the notion of the Hodge Star as a local operator.

Let $x \in \mathbb{R}^{n\times d}$ be a $k$-form on some mesh element, where $n$ is the number of $k$-form elements and $d$ is the feature dimension, and $Q, K, V$ are the linearly mapped queries, keys and values respectively. For each $k$-element $i \in \{1, \ldots, n\}$, we define a sparsity pattern $S_i$, which is the set of key positions that the query at position $i$ can attend to. Then, the update of each $k$-form element feature vector $x_{i}$ through the sparse attention mechanism can be formalized as
\begin{equation}
    x_{i} = \sum_{j \in S_i} A_{ij} V_{j}
\end{equation}
where, $A_{ij}$ is the attention weight between elements $i$ and $j$, calculated as $A_{ij} = \underset{j \in S_i}{softmax} (Q_i \cdot K_j^T) / \sqrt{d}$.

To construct the sparsity pattern $S_i$ for each k-form element $i$, we extract local neighborhoods by performing breadth-first search (BFS) on the adjacency structure of the corresponding mesh element. In addition to local connections, we introduce random connections to further enhance the connectivity of the attention graph. Following the approach of \cite{roy2021efficient}, we select $\sqrt{n}$ neighbors for each mesh element, where $n$ is the total number of elements in the mesh.

In practice, the operators $L_v$, $L_e$, and $L_f$ from \cref{eq:learned-operators} (corresponding to vertex, edge, and face operators, respectively) are not explicitly materialized. Instead, each matrix component of these operators is applied sequentially to the input features, from right to left. This approach, combined with the sparsity pattern described above, results in an overall computational complexity of $O(n^{1.5}d)$.

\subsection{Input Features and Positional Embeddings} \label{sec:input-features-and-positional-embeddings}

HodgeFormer input tokens are directly formed from features on vertices, edges, and faces to enable the application of DEC-based formulations. In addition, positional information is directly incorporated in the architecture as \textit{xyz} coordinates through the input features for each mesh element.
Specifically, we consider three types of input features: \textit{(i) coordinates}, \textit{(ii) normals} and \textit{(iii) areas}: The \textbf{vertex features} $x_{v_{in}}$ consist of the point 3D coordinates, vertex normals, $n_{v_i}$ calculated as the weighted average of incident face normals $n_{f_i}$ and the vertex associated cell area, calculated as the weighted average of incident face areas.
The \textbf{edge features} $x_{e_{in}}$ consist of the point 3D coordinates of edge vertices,
and the vertices opposite to the respective edge, edge normals calculated as the average of incident vertex normals, and the lengths of the edges that belong to the incident faces.
The \textbf{face features} $x_{f_{in}}$ consist of the point 3D coordinates of the face vertices ordered according to the face's orientation, the face normal and the face area.
These design choices aim to incorporate information associated both to the primal and the dual mesh. The effect of each type of input features is showcased in the ablation study in \cref{results:variation-experiments-feats}.

\subsection{Embedding Layer} \label{sec:embedding-layer}

The input features mentioned in \cref{sec:input-features-and-positional-embeddings}, $x_{v_{in}}$, $x_{e_{in}}$ and $x_{f_{in}}$ are transformed by an embedding layer into the corresponding matrices $x_v$, $x_e$ and $x_f$. The embedding layer (\cref{eq:neighbor-embedding}), operates separately on each mesh element and involves two steps: (i) Feature aggregation over one-hop neighborhood defined by the mesh element's adjacency structure and (ii) embedding to the target dimension via a feed-forward MLP.
\vspace{-1em}
\begin{align}
    x_{k} &= MLP(x_{k_{in}} + A_{k} \cdot x_{k_{in}}),      &  k \in \{v, e, f\}
    \label{eq:neighbor-embedding}
\end{align}
By performing one-hop feature aggregation, we allow our network to take into account neighbors' information before embedding the input features to the target dimension.

\begin{table*}[!htbp]
    \centering
    \begin{tabular}{@{}p{3.4cm}clp{1.0cm}lp{1.0cm}lp{1.2cm}p{1.2cm}p{1.2cm}@{}}
    \toprule   
    \textbf{Method} & 
    \textbf{Type}  & 
    \textbf{Acts} & 
    \textbf{Eigen} & 
    \textbf{SHREC11} & 
    \textbf{Cube} &
    \textbf{Human} &  
    \textbf{COSEG} & 
    \textbf{COSEG} & 
    \textbf{COSEG} \\
    & & \textbf{On} &\textbf{Decomp.} & \textbf{(split-10)} & \textbf{Engrav.} & \textbf{Simplified} 
    & \textbf{Vases} 
    & \textbf{Chairs} 
    & \textbf{Aliens} \\

\midrule  
    HodgeNet \cite{smirnov2021hodgenet}         & mlp  & v   & Yes & 94.7\%  & n/a    & 85.0\%  & 90.3\% & 95.7\% & 96.0\%  \\
    DiffusionNet \cite{sharp2020diffusion}      & mlp  & v   & Yes & 99.5\%  & n/a    & 90.8\%  & n/a     & n/a     & n/a   \\
    LaplacianNet \cite{qiao1910laplaciannet}    & mlp  & v   & Yes & n/a     & n/a    & n/a     & 92.2\% & 94.2\% & 93.9\%  \\
    Laplacian2Mesh \cite{dong2023laplacian2mesh}& cnn  & v   & Yes & \textbf{100.0\%} & \textbf{91.5\%} & \textbf{88.6\%}  & 94.6\% & 96.6\% & 95.0\%  \\
    MeT \cite{vecchio2023met}                   & trns & f   & Yes &  n/a    & n/a    & n/a     & \textbf{99.8\%} & \textbf{98.9\%} & \textbf{99.3\%}  \\
    \hline
    MeshCNN \cite{hanocka2019meshcnn}           & cnn  & e   & No  & 91.0\%  & 92.2\% & 85.4\%  & 92.4\% & 93.0\% & 96.3\%  \\
    PD-MeshNet \cite{milano2020primal}          & cnn  & ef  & No  & 99.1\%  & 94.4\% & 85.6\%  & 95.4\% & 97.2\% & 98.2\%  \\
    MeshWalker \cite{lahav2020meshwalker}       & rnn  & v   & No  & 97.1\%  & 98.6\% & n/a     & \textbf{99.6\%} & 98.7\% & \textbf{99.1\%}  \\
    SubDivNet \cite{hu2022subdivision}          & cnn  & f   & No  & 99.5\%  & \textbf{98.9\%} & \textbf{91.7\%}  & 96.7\% & 96.7\% & 97.3\%  \\
    EMNN (MC+H) \cite{trang20243}                 & gnn & ef & No & \textbf{100\%}$^\dagger$ & n/a & 88.7\%$^\dagger$  & n/a & n/a & n/a  \\
    EGNN (MC+H) \cite{trang20243} & gnn & ef & No & 99.6\%$^\dagger$ & n/a & 
    87.2\%$^\dagger$ & n/a & n/a & n/a  \\
    \textbf{HodgeFormer (ours)}                 & trns & vef & No & 98.7\% & 95.3\% & 90.3\%  & 94.3\% & \textbf{98.8\%} & 98.3\%  \\
\bottomrule
\end{tabular}
\vspace{-0.5em}
    \caption{Performance of different methods on mesh classification and mesh segmentation tasks. Mesh classification is evaluated on the 30-class SHREC11\cite{lian2011shrec} dataset evaluated on splits of 10 samples per class (split-10) and the Cube Engraving dataset \cite{hanocka2019meshcnn}. Mesh segmentation is evaluated on the Human dataset in its simplified version of \citet{milano2020primal} with hard ground truth labels at faces, as well as on the Shape COSEG dataset for the categories of Vases, Chairs and Aliens. HodgeFormer achieves results comparable to the state-of-the-art without spectral features, eigenvalue decomposition operations or complex complementary structures. We report their base architecture, the mesh elements they operate on (\textit{v}: vertices, \textit{e}: edges, \textit{f}: faces) and whether they depend on eigen-decomposition methods. The abbreviation "trns" denotes the transformer architecture. Notes: The EMNN (MC+H) and EGNN (MC+H) results shown in the table are as reported by \cite{trang20243}.\\
    $^\dagger$ clarifies that while EGNN and EMNN theoretically do not need pre-calculated Heat Kernel Signatures (HKS), the authors report precomputation of HKS values for both the SHREC and Human Segmentation benchmark datasets \cite[Sec.5,p.5]{trang20243}.
    }
    \label{table:results}
        \vspace{-1.0em}
\end{table*}

\section{Experimental Evaluation}
\label{sec:evaluation}

\subsection{Evaluation Setup}
\label{sec:evaluation-setup}

The HodgeFormer architecture is implemented in PyTorch \cite{paszke2019pytorch} with standard backpropagation. The BFS operations for local neighborhood extraction are based on sparse matrix operations using the \textit{graphblas} framework \cite{graphblas_19}, \cite{graphblas_22}. All experiments and measurements are performed on a single Nvidia RTX 4090 GPU with 24GB VRAM, paired with Intel i9-14900KF processors and 64GB of system RAM. 

In all experiments, we use a common latent dimension $d=256$ across vertices, edges, and faces as well as in the attention mechanism for the $W_Q, W_K, W_V$ mappings. The MLP hidden layers have a dimensionality of $d_h=512$. For the multi-head attention mechanisms, we use $h=4$ heads throughout the architecture, with each head operating on $d_k = d/h = 64$-dimensional space. 

HodgeFormer layers use sparse attention, where each input element attends over $\sqrt{n}$ elements, consisting of local neighbors and random connections in a $4:1$ ratio. Plain Transformer Layers follow a standard implementation except for the incorporation of a linear attention mechanism as in ~\cite{katharopoulos2020transformers}. All MLP layers use \textit{ReLU} non-linearities where required. We adopt the pre-LN architecture variant with residual connections, placing Layer Normalization (LayerNorm) before the input to each layer block. Following \cite{roy2021efficient}, we remove scale and bias terms from LayerNorm. Dropout is employed as a regularization mechanism both in the HodgeFormer and the Transformer layers. Finally, in all experiments we employ the neighbor-aware embedding layer as described in \cref{sec:embedding-layer}.

\subsection{Experiments and Results}
\label{sec:evaluation-results}

The HodgeFormer architecture achieves competitive results with state-of-the-art models in common tasks across multiple datasets without utilizing spectral features, eigenvalue decomposition operations, or complementary structures.

The architecture is validated with experiments on the tasks of mesh classification and mesh segmentation using benchmark datasets. For all experiments, input meshes are zero-centered and scaled to the unit sphere, during both training and testing. During training, we apply random rotations and small perturbations to vertex positions along mesh edges for data augmentations. 
\textcolor{annotation}{All reported results are obtained using architectures of $N=6$ HodgeFormer layers mixed with Transformer layers, in either a $2:1$ or $6:2$ ratio.}
Optimization is performed via standard backpropagation using the Adam optimizer, with learning rates decayed using a cosine annealing strategy. The following hyperparameters were tuned to each problem: learning rate in the range $[10^{-3}, 10^{-4}]$, batch size in $[4, 16]$ and number of training epochs in $[200, 300]$.

The task of mesh classification is evaluated on the SHREC-11\cite{lian2011shrec} and Cube Engraving \cite{hanocka2019meshcnn} datasets. SHREC-11 is a dataset of 600 meshes representing 30 categories of 20 shapes each. Following prior work, we train on randomly selected splits of 10 samples per class and evaluate on the remaining data. Cube Engraving is a synthetic dataset of 2D shapes engraved on a randomly chosen face of a cube, comprising 4,381 shapes from 22 distinct categories. In both experiments, we optimize against a cross-entropy loss with a label smoothing factor of $0.2$. HodgeFormer achieves competitive results compared to other methods (\cref{table:results}). For SHREC-11, the best-performing model uses 6 HodgeFormer layers followed by 2 Transformer layers operating on all mesh elements. In contrast, for Cube Engraving, the top model also uses 6 HodgeFormer layers followed by 2 Transformer layers, but operates only on vertex features.

For the mesh segmentation task, our goal is to predict labels for every face of a mesh. We evaluate HodgeFormer architecture on four datasets: on the \textit{Human-part-segmentation} dataset in its simplified version \cite{milano2020primal} with hard ground truth labels at faces, and the \textit{Vases}, \textit{Chairs}, and \textit{Aliens} categories from the \textit{Shape COSEG} dataset. In all experiments, we optimize against a class-weighted cross-entropy loss with an additional label smoothing factor of $0.2$. For each dataset, we report the highest performing variant of our architecture. \textcolor{annotation}{For the \textit{Human} dataset, our highest performing model was a $6$-layered HodgeFormer model mixed with 2 Transformer layers in a $2:1$ ratio, as presented in Fig. S6a, whereas for \textit{Vases}, \textit{Chairs} and \textit{Aliens} datasets, our highest performing model was a $6$-layered HodgeFormer model followed by two Transformer layers, as presented in Fig. S6b.}
As \cref{table:results} demonstrates, our approach performs comparatively to state of the art without spectral features, eigenvalue decomposition operations, or complex complementary structures.

\subsection{Model Variations Study}

To evaluate the importance of different architecture components and parameter choices, we varied our base model in different ways and measured its performance on a segmentation task. Specifically, in \cref{results:variation-experiments-embed}, we compare the main architecture layers \textit{HodgeFormer} and a vanilla \textit{Transformer} layer along with different embedding layers, i.e., the \textit{feed-forward} and the \textit{neighbor} embedding layers.
In \cref{results:variation-experiments-feats}, we test combinations of different input features based on their type by distinguishing between \textit{coordinates}, \textit{normals}, and \textit{areas}. In Table ~\ref{results:variation-experiments-nbors}, we evaluate the effect of the number of neighbors taken into account by the sparse attention component on the performance of the HodgeFormer layers. 
For all experiments, we consider an architecture of $4$ layers acting only on mesh vertices with embedding dimension $256$ and hidden dimension $512$, and a learning rate of $5\mathrm{e}{-4}$, and we report test accuracy results, averaged over $5$ experiment trials with random training splits. We also consider $\sqrt{n}$ neighbors for each set of $n$ mesh elements with the exception of the experiment of \cref{results:variation-experiments-nbors}. 

In \cref{results:variation-experiments-embed}, the HodgeFormer layers clearly outperform vanilla Transformer layers. At the same time, embedding input features while considering their neighbors offers an edge in initial training steps and results in an overall better performance. In \cref{results:variation-experiments-nbors}, we observe the effect of the number of neighbors for vertices and edges on the HodgeFormer performance. \textcolor{annotation}{
Let $s_k$ be the number of neighbors each element of a $k$-form attends to for $k \in \{v, e, f\}$. 
For $s_v=1, s_e = 1$, the performance drops significantly as the attention mechanism takes into account only self-contributions for vertices and edges.
For $s_v=32, s_e=48$, which approximates $\sqrt{n}$ for meshes in the COSEG vases dataset, the performance is high. Yet, for larger numbers of neighbors, we get diminishing returns in model performance.}
In \cref{results:variation-experiments-feats}, we observe that the HodgeFormer layer is capable of achieving satisfactory results using simple coordinate features, although exhibiting lower performance and slower convergence. 
However, using normals as the sole feature set proves insufficient, resulting in models with poor performance and limited generalization capability.

\begin{figure*}[t]
    \begin{center}    
        \centerline{
            \includegraphics[width=0.7\linewidth]{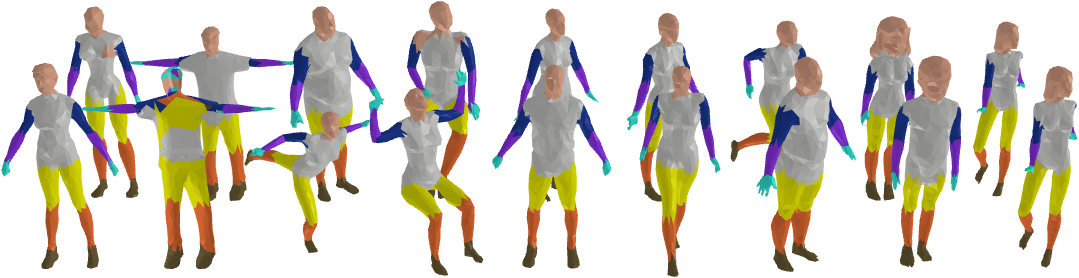}
        }
        \vspace{-1.0em}
        \caption{Mesh segmentation results on the Human Body test set.}
        \label{fig:human-segmentation-results}
    \end{center}
    \vspace{-3.0em}
\end{figure*}

\subsection{Efficiency}

For the HodgeFormer architecture, preprocessing involves data loading, augmentation, and BFS operations for local neighborhood extraction (enabling sparse attention), all performed on-the-fly during data loading without separate precomputation, supporting continuous data streaming across multiple cores.
\cref{table:hodgeformer_performance_exps} compares the runtime performance of HodgeFormer with other recent methods on the human segmentation task, under typical training and testing configurations. HodgeFormer has an attractive runtime profile compared to other methods with fast training, high GPU utilization and no precomputation steps, with a significant part of its execution time spent in I/O and preprocessing, mostly in the BFS-based neighborhood extraction.

\begin{table}[!htbp]
    \centering
    \begin{tabular}{@{}p{3.3cm}p{2.9cm}r@{}}
\toprule
    \textbf{Transformer} & \textbf{Embedding} & \textbf{Vases} \\ 
\midrule                   
    Vanilla Transformer & FNN                       & 84.82\%                \\ 
    Vanilla Transformer & Neighbors + FNN           & 87.28\%                \\ 
    HodgeFormer         & FNN                       & 91.67\%                \\ 
    HodgeFormer         & Neighbors + FNN           & 92.02\%                \\ 
\bottomrule
\end{tabular}
\vspace{-0.5em}
    \caption{Ablation study on the effect of the Neighbor Embedding layer (\cref{sec:embedding-layer}) evaluated on the COSEG Vases dataset. FNN refers to simple Feed-Forward Neural Network.
    }
    \label{results:variation-experiments-embed}
    \vspace{-1.5em}
\end{table}

\begin{table}[!htbp]
    \centering
    \begin{tabular}{@{}p{6.7cm}r@{}}
\toprule
    \textbf{Input Features}    & \textbf{Vases} \\ 
\midrule                         
    coords                     & 86.57\%                \\
    normals                    & 56.41\%                \\
    coords-normals             & 91.66\%                \\
    coords-areas               & 88.22\%                \\
    all (coords-normals-areas) & 92.02\%                \\
\bottomrule
\end{tabular}
\vspace{-0.5em}
    \caption{Comparing input feature effect on HodgeFormer architecture performance, evaluated on the COSEG Vases dataset.}
    \label{results:variation-experiments-feats}
    \vspace{-1.5em}
\end{table}

\begin{table}[!htbp]
    \centering
    \begin{tabular}{@{}p{6.7cm}r@{}}
\toprule
    \textbf{No. Neighbors (\# V, \# E)}    & \textbf{Vases} \\ 
\midrule
    1v, 1e                 & 88.37\%                \\
    8v, 12e                & 90.72\%                \\
    16v, 24e               & 90.84\%                \\
    32v, 48e               & 92.02\%                \\
    64v, 96e               & 92.06\%                \\
    128v, 196e             & 92.13\%                \\
\bottomrule
\end{tabular}
\vspace{-0.5em}
    \caption{Ablation study on number of neighbors for the HodgeFormer Sparse Attention, evaluated on the COSEG Vases dataset.}
    \vspace{-1em}
    \label{results:variation-experiments-nbors}
\end{table}

\begin{table}
    \centering
    \begin{tabular}{@{}p{2.7cm}p{1.0cm}p{1.0cm}p{1.0cm}p{0.8cm}@{}}
\toprule
        &  Exec & GPU & GPU   & Peak    \\
        & Time (s)     & Time  (s)    & Batch (ms)     & Mem (GB) \\
\midrule
    \multicolumn{5}{c}{Training (batch size $=12$)} \\
\midrule
    MeshCNN               &  26.84  & 25.59  & 806.29  &  4.41  \\
    Laplacian2Mesh        & 641.34  & 605.17 & 423.57  &  4.80  \\
    \textbf{HodgeFormer}  &  17.58  & 8.36   & 263.57  & 14.21  \\
\midrule
    \multicolumn{5}{c}{Testing (batch size $=1$)} \\
\midrule
    MeshCNN \cite{hanocka2019meshcnn}               &  1.36   & 1.09   & 60.40   & 0.19  \\
    Laplacian2Mesh\cite{li2022laplacian}        &  2.96   & 0.13   &  7.98   & 0.95  \\
    \textbf{HodgeFormer}  &  1.97   & 0.40   & 22.41   & 0.41  \\
\bottomrule
\end{tabular}
\vspace{-0.6em}
    \caption{Runtime performance comparison of HodgeFormer and other mesh-based methods on train and test sets of Human segmentation dataset, under typical training and testing configurations. We report the following metrics: (a) Execution time per epoch, (b) GPU time per epoch, (c) GPU time per batch, and (d) Peak GPU memory usage. All reported times are averaged over multiple epochs or batches, as appropriate.
    }
\label{table:hodgeformer_performance_exps}
\vspace{-1.5em}
\end{table}

\section{Discussion, limitations and conclusion}
\label{sec:conclusion}
In this work, we draw connections between discrete Hodge operators and Transformer-style attention mechanisms and propose a general architecture for deep learning on triangular meshes named HodgeFormer. Our architecture omits the construction of commonly used techniques such as expensive spectral features, eigenvalue decomposition operations, and complex complementary structures.

Despite these omissions, HodgeFormer retains its expressiveness while fully leveraging the hardware-friendly properties of the Transformer architecture. In addition, by employing sparse attention mechanisms, the HodgeFormer architecture results in an overall complexity of $O(n^{1.5}d)$. HodgeFormer was tested on classification and segmentation tasks, demonstrating performance comparable to state-of-the-art methods.
Strategies such as patching in MeshMAE~\cite{liang2022meshmae} or mesh simplification preprocessing in MeshWalker~\cite{lahav2020meshwalker} are part of future research on large-scale meshes and large-scale unsupervised training. 

Compared to a standard Transformer, our architecture has higher memory requirements due to the use of multiple attention mechanisms per layer. This requirement is partially mitigated by the employment of sparse attention. Although the proposed architecture has low complexity and is fully parallelizable, its implementation relies heavily on gather operations. Investigation of preprocessing strategies for input meshes with reordering algorithms such as the Cuthill-McKee \cite{cuthill1969reducing} may allow the use of contiguous operations enabling better hardware utilization.
\section*{Acknowledgements}
\label{sec:acknowledgements}

This research was supported by TUM Innovation Network CoConstruct N2201 and TUM Georg Nemetschek Institute - Artificial Intelligence for the Built World.

\clearpage 

\section*{Supplementary material}
\vspace{1em}
\appendix 
\setcounter{equation}{0}
\setcounter{table}{0}
\setcounter{figure}{0}
\renewcommand\thefigure{S\arabic{figure}}
\renewcommand\thetable{S\arabic{table}}
\renewcommand\theequation{S\arabic{equation}}

\section{End-to-end Architecture}
\subsection{Multi-head Hodge Attention on vertices, edges and faces}

This section presents a detailed analysis of the HodgeFormer attention mechanism, demonstrating how the multi-head Hodge Attention component operates across different mesh elements. By combining the Hodge star operator formulation (Eq.~11) with the attention-based operators (Eq.~5), the Hodge Laplacians $L_v$, $L_e$, and $L_f$ for vertices (0-forms), edges (1-forms), and faces (2-forms) are reformulated as the attention-based expressions given in equations~\ref{eq:supp_Lv}, \ref{eq:supp_Le}, and \ref{eq:supp_Lf}, respectively.

\begin{align} 
    L_v \enspace := \enspace & \, \star_0^{-1}(x_v) \cdot d_0^T \cdot \star_1(x_e) \cdot d_0 \notag \\
    \label{eq:supp_Lv}
   := \enspace & \sigma\left({\frac{Q_v K_{v}^T}{\sqrt{d_h}}}\right) \cdot d_0^T  \cdot  \sigma\left({\frac{Q_e K_{e}^T}{\sqrt{d_h}}}\right) \\
L_e \enspace := \enspace & \, d_0 \cdot \star_0^{-1}(x_v) \cdot d_0^T \cdot \star_1(x_e) \, +  \notag \\ 
    & \quad \, \, \, \,          \star_1^{-1}(x_e) \cdot d_1^T  \cdot \star_2(x_f) \cdot d_1 \notag \\
    \label{eq:supp_Le}
:= \enspace & \, d_0 \cdot \sigma\left({\frac{Q_v K_{v}^T}{\sqrt{d_h}}}\right) \cdot d_0^T \cdot \sigma\left({\frac{Q_{e1} K_{e1}^T}{\sqrt{d_h}}}\right) \, + \notag \\ 
    & \quad \, \, \, \,          \sigma\left({\frac{Q_{e2} K_{e2}^T}{\sqrt{d_h}}}\right) \cdot d_1^T  \cdot \sigma\left({\frac{Q_f K_{f}^T}{\sqrt{d_h}}}\right) \cdot d_1 \\
    L_f \enspace := \enspace & \, d_1 \cdot \star_1^{-1}(x_e) \cdot d_1^T \cdot \star_2(x_f) \notag \\
    \label{eq:supp_Lf}
    := \enspace & \, d_1 \cdot \sigma\left({\frac{Q_{e2} K_{e2}^T}{\sqrt{d_h}}}\right) \cdot d_1^T \cdot \sigma\left({\frac{Q_{f} K_{f}^T}{\sqrt{d_h}}}\right)
\end{align}

\noindent The updated features for each mesh element are computed by applying the respective Hodge Laplacian to the corresponding value vectors:
\begin{align}
    \label{eq:supp_xv}
    x_v &= L_v \cdot V_v \quad \text{(vertex features)} \\
    \label{eq:supp_xe}
    x_e &= L_e \cdot V_e \quad \text{(edge features)} \\
    \label{eq:supp_xf}
    x_f &= L_f \cdot V_f \quad \text{(face features)}
\end{align}

\begin{figure}[ht]
    \begin{center}
\centerline{\includegraphics[width=0.98\linewidth]{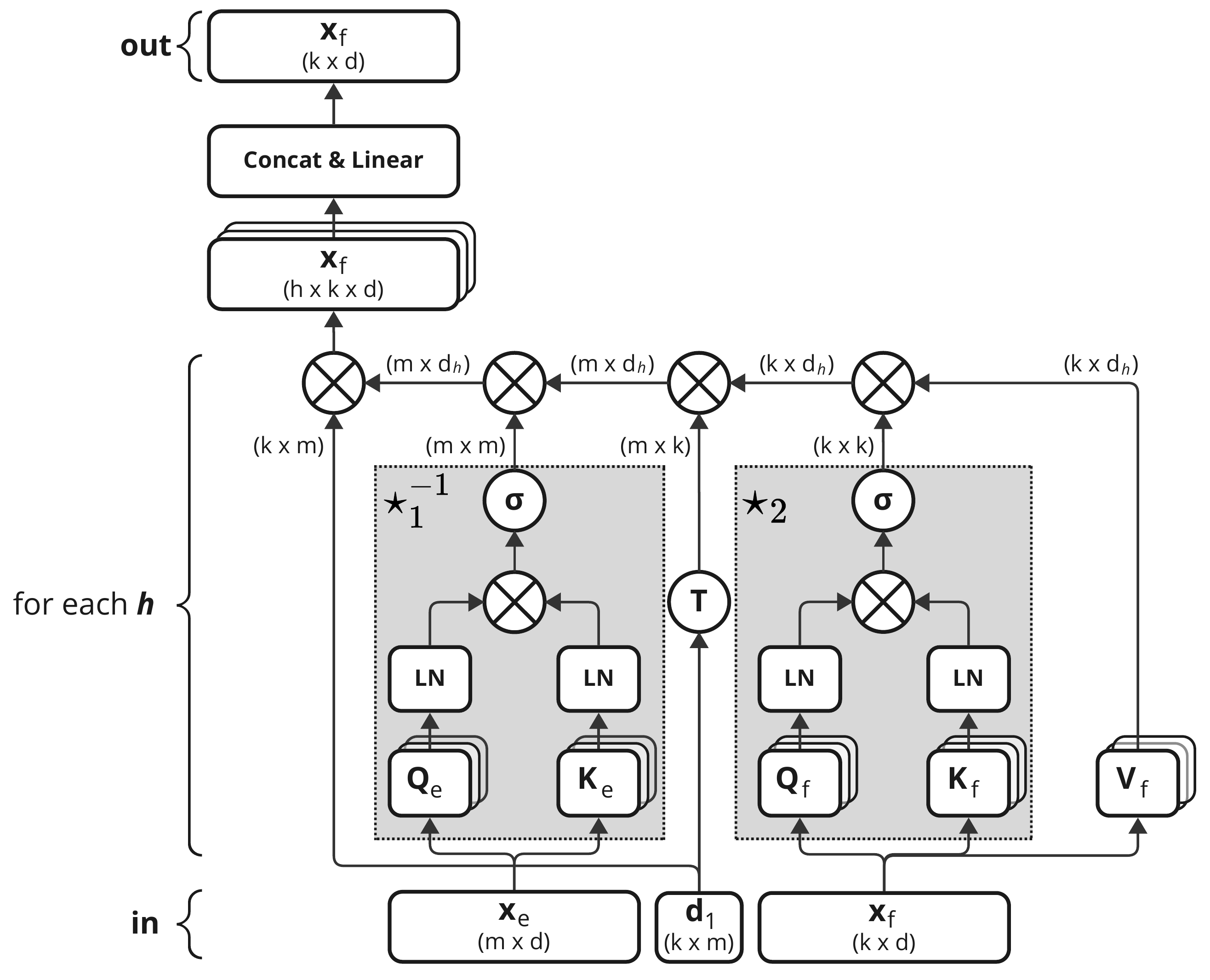}}
        \caption{Multi-head Hodge Attention applied to latent face features $x_f$. The multi-head attention mechanism learns data-driven Hodge Star matrices $\star_1^{-1}$ and $\star_2$.}
        \label{fig:formation_xf}
    \end{center}
\end{figure}

\begin{figure}[ht]
    \begin{center}
\centerline{\includegraphics[width=0.98\linewidth]{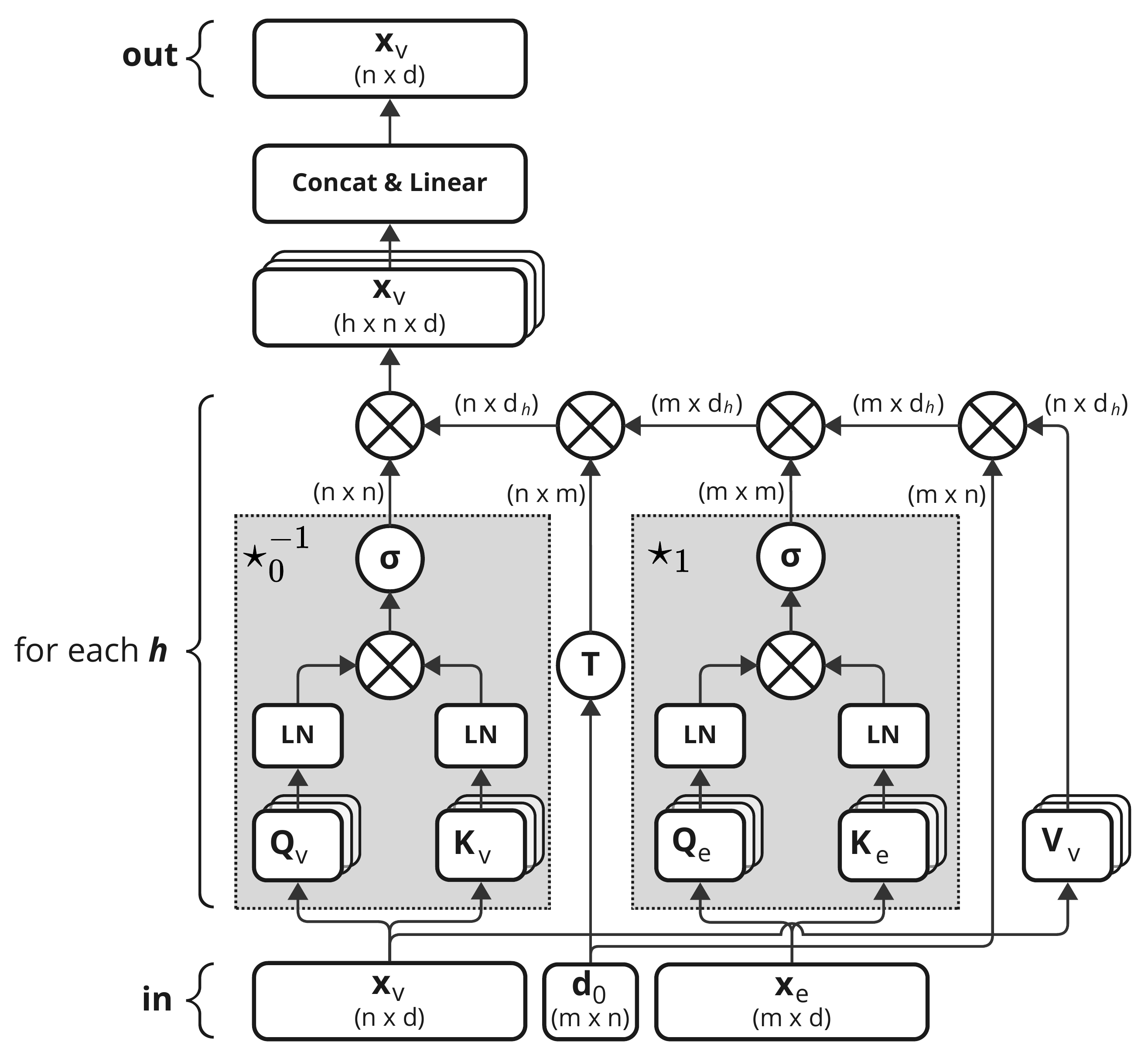}}
        \caption{Multi-head Hodge Attention applied to latent vertex features $x_v$. The multi-head attention mechanism learns data-driven Hodge Star matrices $\star_0^{-1}$ and $\star_1$.}
        \label{fig:formation_xv_2}
    \end{center}
\end{figure}

\begin{figure*}[ht]
    \begin{center}
\centerline{\includegraphics[width=0.99\linewidth]{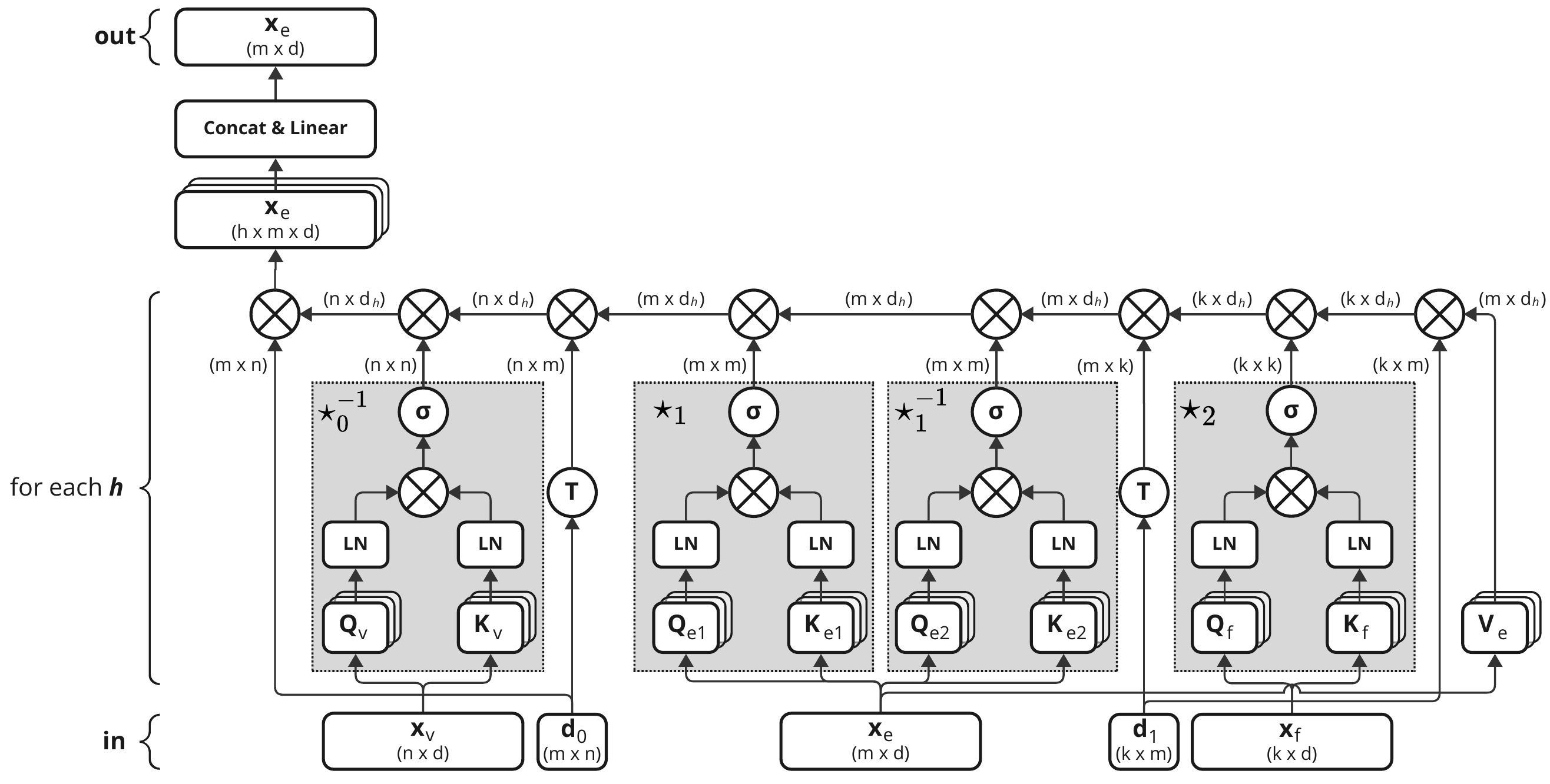}}
        \caption{Multi-head Hodge Attention applied to latent edge features $x_e$. The multi-head attention mechanism learns data-driven Hodge Star matrices $\star_0^{-1}$ and $\star_1$.}
        \label{fig:formation_xe}
    \end{center}
    \vspace{-2em}
\end{figure*}

This formulation ensures that the attention mechanism respects the topological structure of the mesh while enabling information flow between different dimensional elements (vertices, edges, faces).
The formation of these updated features is illustrated in Figures~\ref{fig:formation_xv_2}, \ref{fig:formation_xe}, and \ref{fig:formation_xf}, which visualize the computational flow described in equations~(\ref{eq:supp_xv}), (\ref{eq:supp_xe}), and (\ref{eq:supp_xf}), respectively.

\subsection{Computational complexity and memory requirements}

\noindent Consider a HodgeFormer layer operating on vertex features $x_v$ by applying the operator $L_v := \star_0^{-1}(x_v) \cdot  d_0^T \cdot \star_1(x_e) \cdot d_0$. Also, let $x_v$ and $x_e$ have dimensions $(n_v,d)$ and $(n_e,d)$ respectively, and assume that for practical applications $n_e\approx3n_v$ via Euler's formula on meshes. The layer performs:
\begin{enumerate}[label=(\alph*)]
\item Application of corresponding maps $W_Q$, $W_K$ and $W_V$ on $x_v$ and $x_e$ with complexity $O(n_v \cdot d^2)$ and $O(n_e\cdot d^2)$.
\item Sparse-dense matrix multiplication of $V_v$ with $d_0$ and $d_0^T$ of dimension $(n_e, n_v)$ with complexity $O(n_ed)$ where $n_e$ is the number of nonzero entries of $d_0$.
\item Computation of Hodge matrices $\star_0^{-1}(x_v)$ and $\star_1(x_e)$, based on the product $QK^T$ between elements of dimensions $(n,d)$ and $(\sqrt{n},d)$ with complexity $O(n_v^{1.5} \cdot d)$ and $O(n_e^{1.5} \cdot d)$ respectively. The operation is performed via $\text{gather}$ operations and the full matrix is not materialized.
\item Multiplication of Hodge matrices with feature vector $V_v$ results in complexity $O(n_v^{1.5} \cdot d)$ and $O(n_e^{1.5} \cdot d)$.
\end{enumerate}
The resulting complexity is $O(nd^2) + O(n^{1.5}d)$ with $d$ fixed and usually much smaller than $n$. In comparison, the eigendecomposition of a sparse $(n, n)$ matrix has complexity $O(kn^2)$ for calculating the first $k$ eigenvectors. This holds for Laplacian positional embeddings as well as for spectral features such as Heat Kernel Signature (HKS).
\noindent \cref{table:hodgeformer_memory_time_reqs} presents experiments evaluating the computational and memory requirements. 

\begin{table}[ht]
    \centering
    \begin{tabular}{@{}lp{1.0cm}p{0.9cm}p{0.9cm}p{0.9cm}@{}}
\toprule
    \textbf{Mesh Size ($n_v$)} & 
    $2^{8}$  & 
    $2^{10}$ &
    $2^{12}$ &
    $2^{14}$  \\
\midrule
    \multicolumn{5}{c}{Compute Time (ms)} \\
\midrule
    \textbf{HF Encoder (Train)}   &  5.78   & 8.98   & 29.72  & 197.9  \\
    \textbf{HF Encoder (Infer)}   &  2.50   & 3.42   & 10.96  & 66.13  \\
    \textbf{HF Layer (Infer)}     &  1.08   & 1.91   & 9.42   & 55.25  \\
\midrule
    \multicolumn{5}{c}{Peak Memory Usage (GBs)} \\
\midrule
    \textbf{HF Encoder (Train)}   &  0.12   & 0.40   & 2.54  & 19.23  \\
    \textbf{HF Encoder (Infer)}   &  0.09   & 0.31   & 2.08  & 15.89  \\
    \textbf{HF Layer (Infer)}     &  0.09   & 0.31   & 2.08  & 15.89  \\
\bottomrule
\end{tabular}
\vspace{-0.6em}
    \caption{Metrics for different mesh sizes with respect to the number of vertices, for a 1-layer HodgeFormer (HF) end-to-end architecture (training and inference) as well as a standalone HodgeFormer layer (inference). The model operates on vertices with a latent embedding dimension $d=256$ and hidden MLP dimension of $d_h=512$. Measured on an Nvidia RTX 4090 GPU.
    }
\label{table:hodgeformer_memory_time_reqs}
\vspace{-0.5em}
\end{table}

\section{Results}

\subsection{Robustness Analysis}

To further evaluate the robustness of the HodgeFormer architecture, we perform an analysis, where we evaluate on meshes with added noise, different topology, removed triangles as well as on incomplete meshes. Specifically, we produce variants of the \textit{Human} dataset's test set as follows:

\begin{itemize}
    \item \textbf{Gaussian Noise:} We add Gaussian Noise (GN) $\epsilon$, of several levels $\lambda \in \{0.005, 0.01, 0.02\}$ calculated w.r.t. the diagonal of the axis-aligned bounding box of each model, i.e. $\epsilon = \lambda \cdot |BB_{diag}|$. 
    \item \textbf{QEM Remeshing:} We use the models of the original \textit{Human} dataset from \cite{maron2017convolutional} and remesh them using the Quadratic Error Metric (QEM) to different target face resolution, i.e. $[1000, 2000]$.
    \item \textbf{Face Removal:} We assign to each face a probability $p$ to be randomly removed, and produce dataset variants with different probabilities in the range $[0.01, 0.20]$.
    \item \textbf{Patch removal:} We assign to each face a probability $p=0.005$ to be randomly selected and remove a large patch of $k$ neighbors where $k \sim \mathcal{U}(8,15)$.
\end{itemize}

For all methods, with the exception of QEM Remeshing, we use the test set of the \textit{Human} dataset. Ground truth for evaluation is defined via nearest-neighbor on the original meshes. We evaluate the pre-trained networks reported in Tab. 5, on the mutated datasets and report the results in \cref{results:robustness-analysis}. The HodgeFormer architecture retains well its performance under noise addition, remeshing, removed faces and added holes.

\begin{table}[!htbp]
    \centering
    \begin{tabular}{@{}p{4.4cm}p{1.6cm}p{1.5cm}@{}}
\toprule
    \textbf{HUMAN Test Set Variant}  & 
    \textbf{Accuracy} & 
    \textbf{Acc.Drop} \\
\midrule                   
    Original                          & 90.3\%  & n/a     \\
    Gaussian Noise ($\lambda=0.005$)  & 88.3\%  & 2.0\%  \\
    Gaussian Noise ($\lambda=0.010$)  & 87.3\%  & 3.0\%  \\
    Gaussian Noise ($\lambda=0.020$)  & 81.1\%  & 9.2\%  \\
    QEM Remesh ($1000F$)              & 87.2\%  & 3.1\%  \\
    QEM Remesh ($2000F$)              & 86.2\%  & 4.1\%  \\
    Face Removal ($p=0.04$)           & 87.8\%  & 2.5\%  \\
    Face Removal ($p=0.10$)           & 85.9\%  & 4.4\%  \\
    Face Removal ($p=0.20$)           & 81.6\%  & 8.7\%  \\
    Patch removal            & 86.8\%  & 3.5\%  \\
\bottomrule
\end{tabular}
\vspace{-0.5em}
    \caption{Robustness evaluation of the HodgeFormer architecture. We report the performance of pre-trained networks on mutated variations of the \textit{Human} dataset test set. The HodgeFormer architecture retains well its performance under noise addition, remeshing, removed faces and added holes.}
    \label{results:robustness-analysis}
    \vspace{-1em}
\end{table}

\begin{figure}[ht]
    \begin{center}    
        \centerline{
            \includegraphics[width=\linewidth]{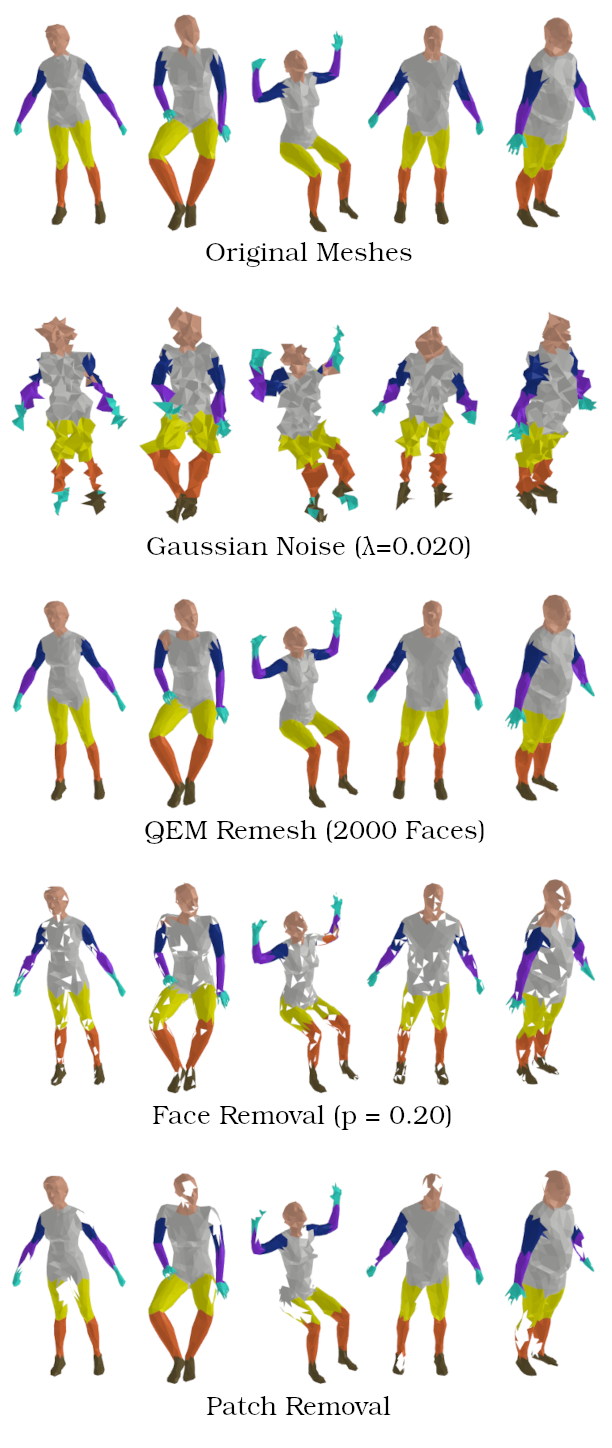}
        }
        \vspace{-1.0em}
        \caption{Mesh segmentation results on selected models from the mutated Human Body test set. The segmentation results remain reasonable and consistent with the initial predictions of the network.}
        \label{fig:robustness-analysis-qualitative-vertical}
    \end{center}
    \vspace{-3.0em}
\end{figure}

\begin{figure}[ht]
    \begin{center}    
        \centerline{
            \includegraphics[width=\linewidth]{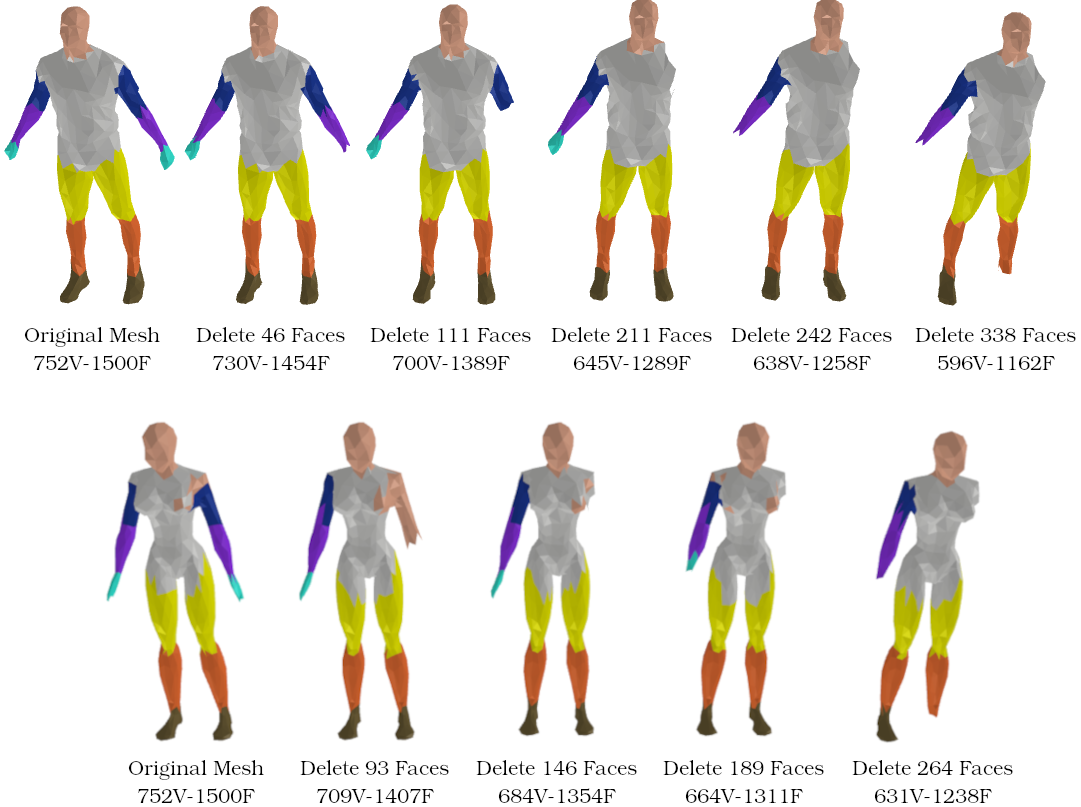}
        }
        \caption{Mesh segmentation results on selected models from the mutated Human Body test set. The segmentation results remain reasonable and consistent with the initial predictions of the network.}
        \label{fig:human-incomplete-segmentation-results}
    \end{center}
    \vspace{-2em}
\end{figure}

\noindent \textbf{Incomplete Meshes:} Additionally, we select two mesh models from the test set of the Human Body Dataset, one with accurate and one with inaccurate segmentation results, and gradually remove body parts from them, similar to \cite{dong2023laplacian2mesh}. Then, we produce segmentations using a pre-trained HodgeFormer network as showcased in \cref{fig:human-incomplete-segmentation-results}. The segmentation results remain reasonable and consistent with the initial predictions of the network on these mesh models, even on the one with inaccurate segmentation.

\subsection{Training on One Sample per Class}

We repeat the experiment performed by \cite{peng_mwformer_2023}, where a model is trained on the 30-class \textit{SHREC11} \cite{lian2011shrec} dataset on splits of only one sample per class (split-1). The model follows the same architecture as Sec.5 with a learning rate of $0.0025$ to compensate for the small dataset size. The results are presented in \cref{tab:split-1} along with (split-10) results for comparison.

\begin{table}[ht]
    \centering
    \begin{tabular}{@{}p{3.5cm}p{2.0cm}p{2.0cm}@{}}
\toprule
    \textbf{Method}   &
    \textbf{SHREC11}  & 
    \textbf{SHREC11}  \\
                      & 
    (split-10)        & 
    (split-1)         \\
\midrule
    SubDivNet \cite{hu2022subdivision} & 99.5\%           & 36.5\%           \\
    MWFormer \cite{peng_mwformer_2023} & 100.0\%          & 41.7\%           \\
    \textbf{HodgeFormer (ours)}        & 98.7\%           & \textbf{45.9\%}  \\
\bottomrule
\end{tabular}
    \caption{Classification performance of different methods on the 30-class \textit{SHREC11}\cite{lian2011shrec} dataset trained on splits of only 1 sample per class (split-1). Results for (split-10) are also reported for comparison.}
    \label{tab:split-1}
\end{table}

\subsection{Variations Study on Mixing Strategies}
\label{sec:mixing-strategies}

This section examines the effect of mixing HodgeFormer and Transformer layers on segmentation performance. Table~\ref{results:variation-experiments-mix} summarizes the performance on the \textit{COSEG} Vases dataset under varying layer compositions. 
We denote $N_H$ as the number of HodgeFormer layers and $R_{H:T}=(N_H:N_T)$ the ratio of the number of HodgeFormer layers to Vanilla Transformer layers.
As shown in Table \ref{results:variation-experiments-mix}, carefully mixing Transformer with HodgeFormer layers has an effect on the final model performance. In general, from the conducted experiments, different mixing strategies would fit different datasets. For the \textit{COSEG} Vases the best result (93.04\%) was obtained by adding two Transformer layers on top of the HodgeFormer layers. Additional layouts of HodgeFormer and Transformer layers are showcased in \cref{sec:hodgeformer-layer-layout}.

\begin{table}[ht]
    \centering
    \begin{tabular}{@{}p{5.5cm}r@{}}
\toprule
    \textbf{Layers} & \textbf{Vases (COSEG)} \\
\midrule
    $N_H=4$ \& $R_{(N_H,N_T)}=4:0$ & 92.02\% \\
    $N_H=4$ \& $R_{(N_H,N_T)}=1:1$ & 92.85\% \\
    $N_H=4$ \& $R_{(N_H,N_T)}=2:1$ & 92.59\% \\
    $N_H=4$ \& $R_{(N_H,N_T)}=4:2$ & 93.04\% \\
    \bottomrule
\end{tabular}
    \caption{Variation experiments comparing different strategies of mixing HodgeFormer with Transformer layers and their effect on the performance, evaluated on the \textit{COSEG} Vases dataset.}
    \label{results:variation-experiments-mix}
\end{table}

\subsection{Variation Study on Label Smoothing}

This section examines the effect of label smoothing in the cross-entropy loss on the model's performance for classification and segmentation tasks. Table~\ref{results:variation-experiments-label-smoothing} reports classification results on \textit{SHREC11} and segmentation results on \textit{COSEG} Vases under different label smoothing values.

\begin{table}[ht]
    \centering
    \begin{tabular}{@{}p{4.6cm}cc@{}}
        \toprule
        \textbf{Label Smoothing} & \textbf{SHREC11} & \textbf{Vases}\\
                                 &  (split-10)      &        (COSEG)\\
        \midrule
            $0.00$ & 97.50\%  & 94.03\% \\
            $0.05$ & 98.33\%  & 94.08\% \\
            $0.10$ & 98.67\%  & 94.12\% \\
            $0.20$ & 98.70\%  & 94.30\% \\
            $0.40$ & 98.33\%  & 93.25\% \\
        \bottomrule
    \end{tabular}
    \caption{Variation study on the effect of label smoothing on the model's performance for classification and segmentation tasks.}
    \label{results:variation-experiments-label-smoothing} 
\end{table}

\subsection{Eigenfunctions of the corresponding learned operator}

\begin{figure}[ht]
    \centering
    \begin{subfigure}[t]{\linewidth}
        \centering
        \includegraphics[width=0.96\linewidth]{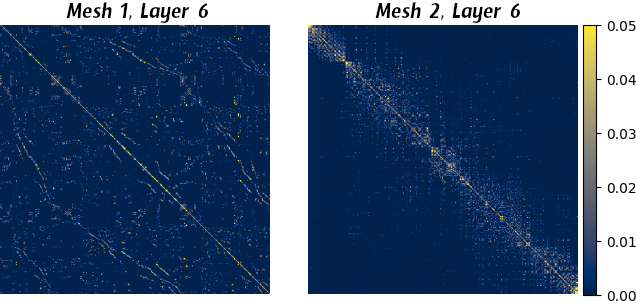}
    \end{subfigure}

    \vspace{0em} 
        \begin{subfigure}[t]{\linewidth}
        \centering
        \includegraphics[width=\linewidth]{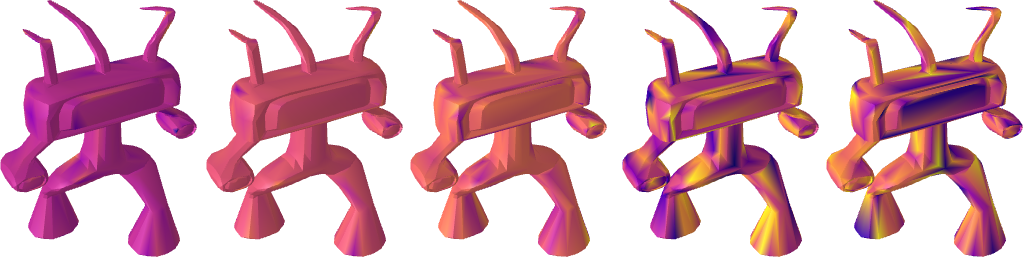}
        \label{fig:bottom-center}
    \end{subfigure}
    \vspace{-2.0em}
        \caption{\footnotesize{Qualitative illustration. Above: Learned attention maps from two samples of the COSEG Aliens dataset. Below: First few eigenfunctions of the corresponding learned operator, $L_6,head_1$ for the 1st sample.}}
    \label{fig:eigenfunctions-attention-maps}
\end{figure}

We compute the eigenfunctions of the corresponding learned operators and visualize them on the mesh geometry, providing insights into how the HodgeFormer layer captures and processes geometric features at different scales and orientations.  Figure \ref{fig:eigenfunctions-attention-maps} at the top, presents the learned attention maps for two sample meshes of \textit{COSEG} Aliens dataset measured in layer 6 of the architecture.
Figure \ref{fig:eigenfunctions-attention-maps} at the bottom presents the first five eigenfunctions of a \textit{COSEG} Aliens dataset sample.

\begin{figure*}[ht]
    \centering
    \includegraphics[width=\linewidth]{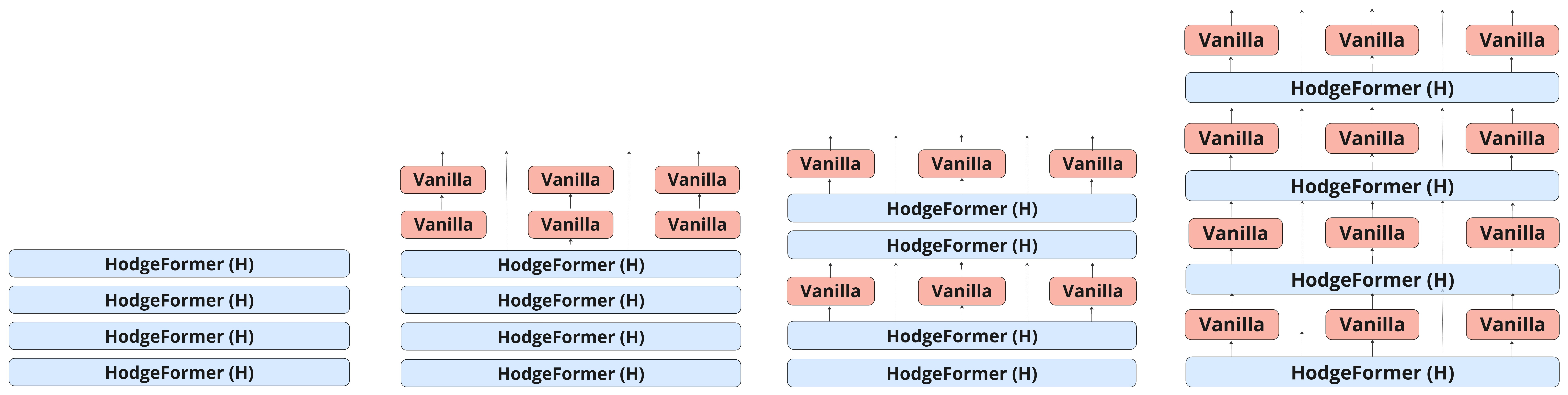}
       
        \begin{subfigure}{0.24\textwidth}
            \caption{}
        \end{subfigure}
    \hfill
        \begin{subfigure}{0.24\textwidth}
        \caption{}
    \end{subfigure}
    \hfill
        \begin{subfigure}{0.24\textwidth}
        \caption{}
    \end{subfigure}
    \hfill
        \begin{subfigure}{0.24\textwidth}
        \caption{}
    \end{subfigure}
\vspace{0em}
     \caption{Visualization of mixing strategies of HodgeFormer and vanilla Transformer layers used in \cref{sec:mixing-strategies}. $N_H$ corresponds to the number of HodgeFormer transformer layers and $R_{(N_H,N_T)}$ corresponds to the ratio of HodgeFormers to vanilla transformers: a) $N_H=4$ \& $R_{(N_H,N_T)}=4:0$, b) $N_H=4$ \& $R_{(N_H,N_T)}=4:2$, c) $N_H=4$ \& $R_{(N_H,N_T)}=2:1$, d) $N_H=4$ \& $R_{(N_H,N_T)}=1:1$ }
     \label{fig:supp_ablation}
\end{figure*}

\subsection{Hodgeformer \& Transformer Layer Layouts}
\label{sec:hodgeformer-layer-layout}

This section includes visualizations of HodgeFormer architecture variations and interleaving strategies, along with notation clarifications. $N_H$ corresponds to the number of HodgeFormer transformer layers and $R_{(N_H,N_T)}$ corresponds to the ratio of HodgeFormers to vanilla transformers.

\begin{figure}[ht]
    \begin{center}   
    
        \begin{subfigure}{\linewidth}
            \centering
            \includegraphics[width=\linewidth]{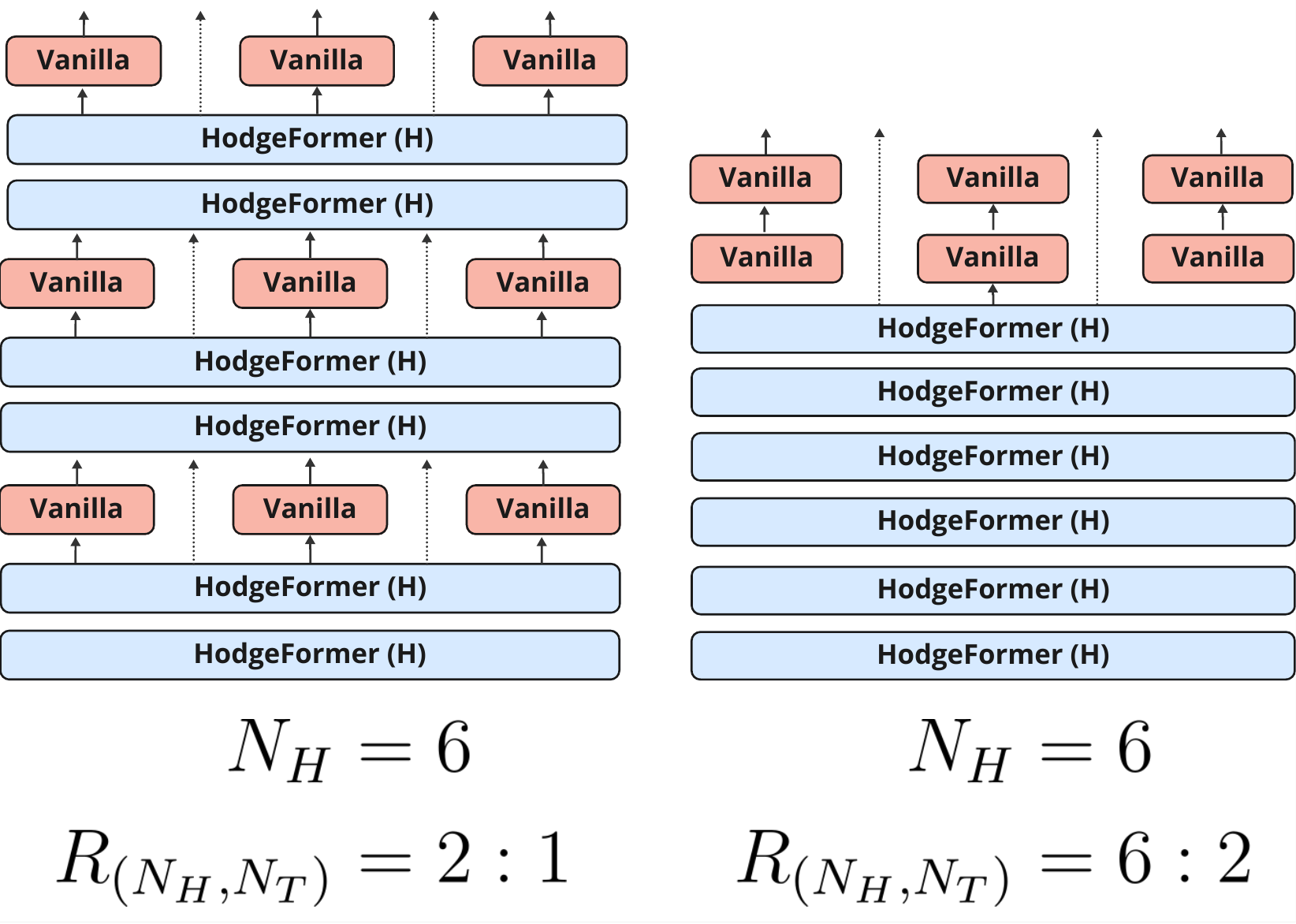}
        \end{subfigure}
    
    \begin{subfigure}{0.48\linewidth}
        \centering
        \caption{}
        \label{fig:HF_variations_a}
    \end{subfigure}
    \hfill 
    \begin{subfigure}{0.48\linewidth}
        \centering
        \caption{}
        \label{fig:HF_variations_b}
    \end{subfigure}

        \caption{HodgeFormer architecture variations. For \textit{Human} dataset, our highest performing model was a $6$-layered HodgeFormer model mixed with 2 Transformer layers in a $2:1$ ratio, as presented in (a), whereas for \textit{Vases}, \textit{Chairs} and \textit{Aliens} datasets, our highest performing model was a $6$-layered HodgeFormer model followed by two Transformer layers,  as shown in (b). }
        \label{fig:hf-variations}
    \end{center}
    \label{fig:HF_variations}
    \vspace{-1em}
\end{figure}


\end{document}